\documentclass[aps,pra,twocolumn,floatfix,reprint]{revtex4-1}
\usepackage{amsfonts}
\usepackage{mathrsfs}	
\usepackage{amsmath}
\usepackage{color}
\usepackage{graphicx}
\usepackage{bm}
\usepackage{amssymb}
\usepackage{xspace}
\usepackage{epstopdf}
\usepackage{dcolumn}
\usepackage{tabularx}
\usepackage{longtable}
\usepackage[colorlinks=true, pdfstartview=FitV, linkcolor=blue, citecolor=blue, urlcolor=blue]{hyperref}
\usepackage[normalem]{ulem}

\begin{document}

\title{Spin-orbit-coupling induced localization in the expansion of an interacting Bose-Einstein condensate}
\author{Chunlei Qu$^{1}$}
\email{chunleiqu@gmail.com}
\author{Lev P. Pitaevskii$^{1,2}$}
\author{Sandro Stringari$^{1}$}
\date{\today }

\affiliation{
$^{1}$INO-CNR BEC Center and Dipartimento di Fisica, Universit\`a di Trento, 38123 Povo, Italy \\
$^{2}$Kapitza Institute for Physical Problems RAS, Kosygina 2, 119334 Moscow, Russia}

\begin{abstract}
By developing a hydrodynamic formalism, we investigate the expansion dynamics of the single-minimum phase of a binary spin-orbit coupled Bose-Einstein condensate, after releasing from an external harmonic trap. 
We find that the expansion of the condensate along the direction of the spin-orbit coupling is dramatically slowed down near the transition between the single-minimum phase and the plane-wave phase. Such a slow expansion, resembling a form of an effective localization, is due to the quenching of the superfluid motion which results in a strong increase of the effective mass. In the single-minimum phase the anisotropic expansion of the Bose gas, which is spin balanced at equilibrium, is accompanied by the emergence of a local spin polarization. Our analytic scaling solutions emerging from hydrodynamic picture are compared with a full numerical simulation based on the coupled Gross-Pitaevskii equations.
\end{abstract}

\maketitle

\section{Introduction}
\label{sec:intro}

Since the first experimental realization of Bose-Einstein condensation in 1995 the expansion of quantum gases, after release of the confining trap, has systematically provided crucial information on the physical properties of such systems. In Ref.~\cite{Anderson1995, Davis1995} the observation of the bimodal expansion at finite temperature actually provided crucial evidence for the  occurrence of Bose-Einstein condensation. Later experiments with larger numbers of atoms and stronger interaction effects have provided strong  evidence for the hydrodynamic nature of the expansion~\cite{Ernst1998,Stenger1998}. The expansion of the condensate has also been employed to characterize the emergence of the superfluid to Mott insulator transition in the presence of a periodic optical lattice~\cite{Greiner2002}, the effects of localization in the presence of disorder~\cite{Roati2008}, the occurrence of Bloch oscillations~\cite{Roati2004} and a large variety of physical phenomena in both Bose and Fermi quantum gases~\cite{Pitaevskii2016}.   

In many cases the condensate is in the so-called Thomas-Fermi limit, where the equation of state of uniform matter can be directly employed in the local density approximation and the mechanism of the expansion is well described using the hydrodynamic formalism of superfluids. In usual superfluids at low temperature, the hydrodynamic picture is based on the irrotationality constraint for the velocity field which takes the most famous expression ${\bf v} =(\hbar/m)\nabla \phi$ where $\phi$ is the phase of the order parameter, $\hbar$ is the reduced Plank constant and $m$ is the atomic mass. The hydrodynamic approach has proven quite useful to describe the macroscopic dynamic behavior of quantum gases, including the study of the low frequency collective oscillations~\cite{Stringari1996} and the expansion of the gas after release of the trap~\cite{Castin1996,Kagan1996,Dalfovo1997}, in excellent agreement with experiments.

The recent experimental investigation of spin-orbit coupled quantum gases has enlarged the horizon of the possible scenarios characterizing the dynamic behavior of quantum gases~\cite{Lin2011,Fu2011,Zhang2012,Ji2015,Wang2012,Cheuk2012,Qu2013,Olson2014} (for recent review articles, see~\cite{Zhai2015,Li2015}). In particular the presence of spin-orbit coupling in binary mixtures of Bose-Einstein condensates is known to affect the irrotationality constraint and the   current-phase relation   
${\bf j} =n (\hbar/m)\nabla \phi$ holding in usual superfluids. This is the consequence of the breaking of Galilean invariance and is at the origin of novel features exhibited by such systems, like the violation of Landau's criterion for superfluidity~\cite{Zhu2012,Zheng2013,Ozawa2012a}, the quenching of the sound velocity~\cite{Martone2012} and of the superfluid density~\cite{Zhang2016}, the emergence of diffused vorticity in the rotational flow~\cite{Stringari2017}. Experimentally a first direct evidence for the violation of the superfluid current-phase relation was given by the quenching of the frequency of the center-of-mass oscillation in harmonically trapped configurations~\cite{Zhang2012} and of the sound velocity~\cite{Ji2015}. In a very recent work \cite{Khamehchi2016} the expansion of a Bose-Einstein condensate in the presence of spin-orbit coupling was investigated with special focus on the   plane-wave phase, where intriguing effects associated with the occurrence of negative values of the effective mass were observed.

\begin{figure}[t!]
\centerline{
\includegraphics[width=8.6cm]{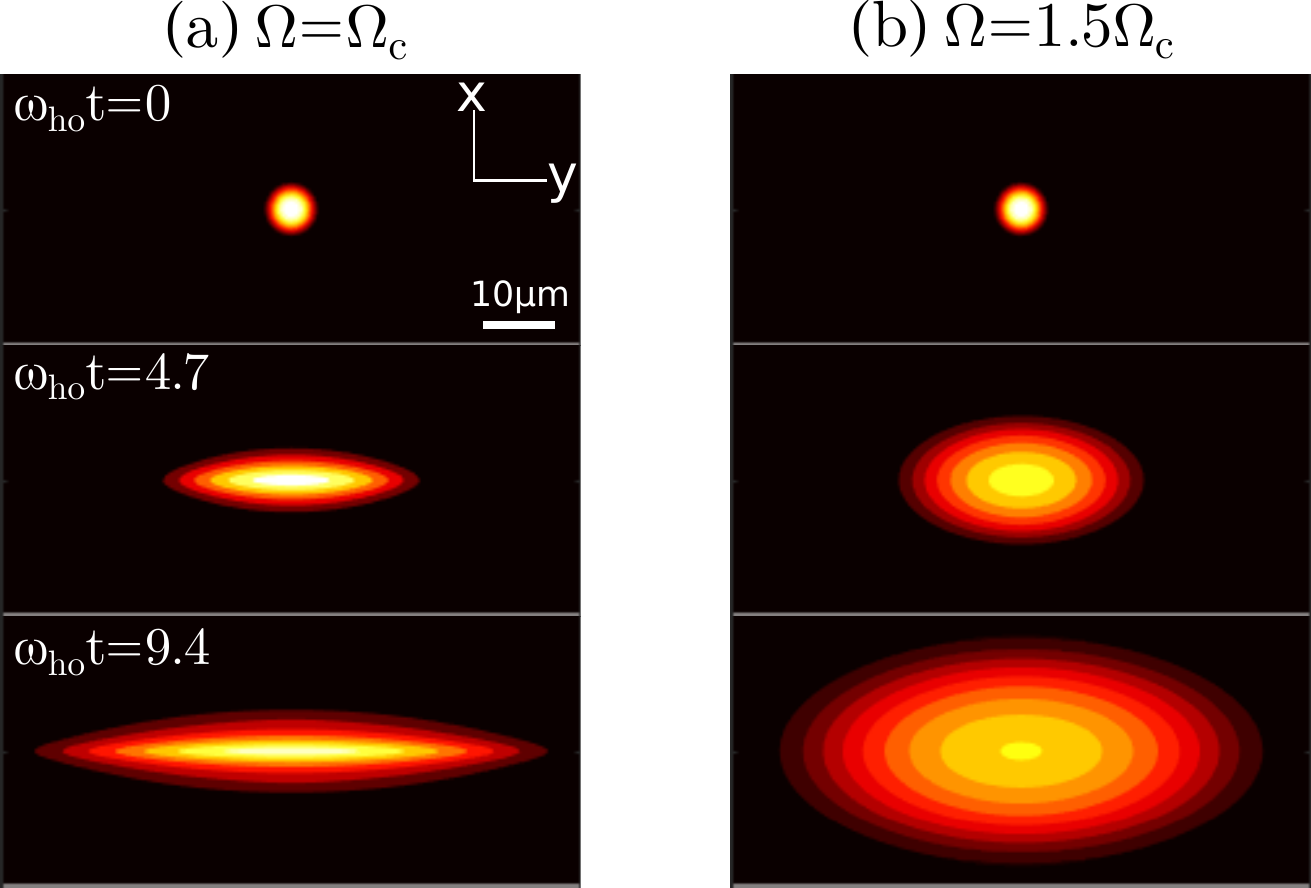}}
\caption{Contour plot of the total integrated density $n(x, y)$ of the expanding spin-orbit coupled condensate for (a) $\Omega=\Omega_\text{c}$ and (b) $\Omega=1.5\Omega_\text{c}$ (In each panel, the brighter the color, the higher the density). For each value of Raman coupling $\Omega$, the density profiles have been shown after an expansion time $\omega_\text{ho}t=0$, $4.7$ and $9.4$, respectively. In the top left panel we also show a length scale corresponding to $10\mu m$. The results are obtained by solving the Gross-Pitaevskii equations with the spin-orbit coupling along the vertical direction. See the main text for the detailed description of the simulation parameters.}
\label{Fig1}
\end{figure}

In the present paper we investigate the  expansion of a spin-orbit coupled Bose gas focusing on the single-minimum phase, where a simplified hydrodynamic description is formulated in terms of scaling variables and analytic solutions to the nonlinear equations characterizing the expansion can be obtained. We find that, for values of the Raman coupling close to phase transition between the single-minimum phase and the plane-wave phase, the expansion along the direction of spin-orbit coupling is dramatically quenched exhibiting an effective localization caused by the lowering of the superfluid flow. A typical example of quenched expansion caused by spin-orbit coupling is shown in Fig.~\ref{Fig1} where the integrated density  $n(x, y)=\int n(\mathbf{r})dz$ obtained by numerically solving the Gross-Pitaevskii equations is reported for different expansion times, starting from an initially isotropic density distribution. At the phase transition, corresponding to the choice $\Omega=\Omega_\text{c}$ for the strength of the Raman coupling, the expansion is frozen along the direction of spin-orbit coupling (the vertical axis).

\section{Hydrodynamic formalism}
\label{sec:hydro}

We consider the following Raman induced spin-orbit coupled single-particle Hamiltonian~\cite{Ho2011,Li2012}
\begin{equation}
H_\text{sp}=\frac{1}{2m}(p_x-\hbar k_0\sigma_z)^2 + \frac{p_y^2}{2m} + \frac{p_z^2}{2m} - \frac{\Omega}{2}\sigma_x
\label{eq:sp}
\end{equation}
where $p_\mu=-i\hbar \nabla_\mu$ ($\mu=x,y,z$) is the canonical momentum, $\hbar k_0$ is the recoil momentum and $\Omega>0$ is the Raman coupling strength. The detuning term $(\delta/2)\sigma_z$ can be set to zero with a proper choice of the frequency of the two Raman lasers. Interaction effects will be taken into account through the interaction energy $V_\text{int}=(1/2)\sum_{\alpha\beta}\int d\mathbf{r}g_{\alpha\beta}n_\alpha n_\beta$ where $n_\alpha$ is the density distribution of the $\alpha$-th component, $g_{\alpha\beta}=4\pi\hbar^2a_{\alpha\beta}/m$ are the coupling constants in different spin channels and $a_{\alpha\beta}$ the corresponding scattering lengths. For simplicity, we assume that the interspin and intraspin interaction constants are the same, i.e., $g_{11}=g_{22}=g_{12}=g$. The spin-orbit coupled Bose-Einstein condensate is known to exhibit a second-order phase transition at the critical Raman coupling strength $\Omega_\text{c}=4E_\text{r}$ where $E_\text{r}=\hbar^2k_0^2/2m$ is the recoil energy. For $\Omega<\Omega_\text{c}$, the lower band of the single particle dispersion exhibits double minima at the wave vectors $\pm k_0\sqrt{1-(\Omega/\Omega_\text{c})^2}$. The condensate picks up one of them spontaneously and the resulting ground state then corresponds to a spin-polarized plane-wave phase. For $\Omega>\Omega_\text{c}$, the lower band has instead only a single minimum at zero momentum and the resulting ground state is named as single-minimum phase or zero-momentum phase which is instead spin balanced at equilibrium.

The spin-orbit Hamiltonian Eq.~(\ref{eq:sp}) is known to affect in a deep way the expression for the current operator along the $x$ direction which takes the form
\begin{equation}
{\hat j}_x=({\hat p}_x -\hbar k_0 {\hat \sigma_z})/m 
\end{equation}
and, as we will explicitly discuss,  causes the violation of the usual current-phase relation ${\bf j} =n (\hbar/m)\nabla \phi$.

As discussed in~\cite{Stringari2017}, the relative phase of the two components is locked for the ground state   and for the low frequency dynamics. Therefore, the dynamic behavior  of the spin-orbit coupled Bose-Einstein condensate is described by the densities of the two components $n_\alpha$ and by their common phase $\phi$. Starting from the coupled Gross-Pitaevskii equations in the presence of spin-orbit coupling   and neglecting quantum pressure effects  one can then derive the following  equations:
\begin{eqnarray}
&& \frac{\partial n}{\partial t} + \frac{\hbar}{m}\nabla\cdot(n\nabla\phi)-\frac{\hbar k_0}{m}\nabla_x s_z=0 \label{eq:continuity} \\
&& \hbar\frac{\partial\phi}{\partial t} + \frac{\hbar^2}{2m}(\nabla\phi)^2 + gn + V_\text{ext}-\frac{\Omega}{2}\frac{n}{\sqrt{n^2-s_z^2}} = 0  \label{eq:Euler} \\
&& -\frac{\hbar^2 k_0}{m}\nabla_x \phi + \frac{\Omega}{2}\frac{s_z}{\sqrt{n^2-s_z^2}} = 0 \label{eq:spinphase}
\end{eqnarray}
where $V_\text{ext}$ is the confining  harmonic   potential,  $n=n_1+n_2$ and $s_z=n_1-n_2$ are the total density and spin density.  Note that these quantities are position- and time-dependent in the expansion dynamics.  From the continuity equation~(\ref{eq:continuity}), it is natural to introduce the superfluid velocity $v_x=(\hbar/m)\nabla_x\phi-\hbar k_0s_z/(mn)$, $v_{y,z}=(\hbar/m)\nabla_{y,z}\phi$. By writing the spin density in terms of the local quasi-momentum $p_x= \hbar\nabla_x \phi$ with the help  of Eq. (\ref{eq:spinphase}) one recovers the hydrodynamic formalism  presented in the recent work~\cite{Khamehchi2016}, based on the inclusion of the lower branch of the single-particle spectrum.

In the following we will consider the single-minimum phase ($\Omega \ge \Omega_\text{c}$), where, at equilibrium, the system is unpolarized and its phase can be chosen equal to zero. To obtain simple analytic expressions, we assume that the spin density accumulated during the expansion is small compared to the total density, i.e., $s_z\ll n$. One can then expand the square roots in the above equations and include terms quadratic in the spin density, generalizing the linearized expressions derived in~\cite{Stringari2017}.  Using Eq.~(\ref{eq:spinphase})  a useful relation between the spin density and the superfluid velocity is then obtained
\begin{equation}
\frac{s_z}{n}=\frac{2\hbar k_0}{\Omega}\frac{m^*}{m}v_x  \label{eq:spindensity2}
\end{equation}
with the effective mass given by 
\begin{equation}
m^*=m\left(1-\Omega_\text{c}/\Omega \right)^{-1}
\end{equation}

Using these relations, after eliminating the spin density, the above differential equations can be recast in the form
\begin{eqnarray}
&& \frac{\partial n}{\partial t} + \nabla\cdot(n\mathbf{v}) = 0 \label{eq:HD1} \\
&& m^*\frac{\partial v_x}{\partial t} + \nabla_x\left(\frac{m^*}{2}v_x^2+gn+V_\text{ext} \right) = 0 \label{eq:HD2} \\
&& m\frac{\partial v_{y,z}}{\partial t} + \nabla_{y,z}\left(\frac{m}{2}v_{y,z}^2+gn+V_\text{ext} \right) = 0
\label{eq:HD3}
\end{eqnarray} 
hereafter called the equations of quadratic hydrodynamics,  where we have retained only quadratic terms in the velocity field. Equations of the same form can also be derived in the plane-wave phase ($\Omega \le \Omega_\text{c}$), where the effective mass is given by $m^*=m(1-(\Omega/\Omega_\text{c})^2)^{-1}$. In the plane-wave phase the equations of quadratic hydrodynamics  have however a more limited range of applicability because in this case  high order terms in the velocity field become soon relevant and cause the appearance of asymmetric effects in the expansion dynamics along the direction of spin-orbit coupling, as shown in \cite{Khamehchi2016}.

\section{Expansion dynamics}
\label{sec:expansion}

Similar to the case in the absence of spin-orbit coupling, the equations~(\ref{eq:HD1})-(\ref{eq:HD3}) of  quadratic hydrodynamics  admit a solution of the inverted parabolic form $n(\mathbf{r}, t)=n_0(t)\left(1-x^2/R_x^2(t)-y^2/R_y^2(t)-z^2/R_z^2(t)\right)$ within the region where $n(\mathbf{r}, t)$ is positive and zero elsewhere~\cite{Castin1996,Kagan1996,Dalfovo1997}. The peak density $n_0(t)$ is determined by the normalization condition $n_0(t)=15N/[8\pi R_x(t)R_y(t)R_z(t)]$ where $N$ is the number of atoms. The Thomas-Fermi radii $R_\mu(t)$ can be written in the scaled form $R_\mu(t)=b_\mu(t)R_\mu^{(0)}$ ($\mu=x,y,z$) in terms of the dimensionless scaling parameters $b_\mu(t)$  and of the Thomas-Fermi radii $R_\mu^{(0)}$ before the expansion. The velocity field is parameterized as $v_\mu(x_\mu, t)=a_\mu(t)x_\mu$. Substituting these ansatz into the hydrodynamic equations, we find $a_\mu(t)=\dot{b}_\mu(t)/b_\mu(t)$  and, after switching off the trapping potential, the equations for the scaling parameters $b_{\mu}$ take the form
\begin{equation}
\ddot{b}_\mu(t)-\frac{\tilde{\omega}_\mu^2}{b_\mu(t)b_x(t)b_y(t)b_z(t)}=0 \; , \label{eq:Db}
\end{equation}
where $\tilde{\omega}_x=\omega_x\sqrt{m/m^*}$ and $\tilde{\omega}_{y,z}=\omega_{y,z}$. The initial conditions are determined by the properties of the system at equilibrium, thus $a_\mu(t=0)=0$ and $b_\mu(t=0)=1$. By solving the above differential equations for $b_\mu(t)$ and $a_\mu(t)$ with these initial conditions, one can obtain the density profile and the velocity field of the condensate   as well as, according to Eq. (\ref{eq:spindensity2}), the spin density during the expansion. It is worth  noticing that, different from the plane-wave phase, the
total density of the single-minimum phase is always symmetric in the spin-orbit coupling direction during the expansion.

Close to the critical transition point from the single-minimum phase to the plane-wave phase, $m/m^*\to 0$ and thus $\tilde{\omega}_x\to 0$. In this regime  it is thus possible to write down the analytical expression~\cite{Castin1996}
\begin{eqnarray}
&& b_{y,z}(\tau) = \sqrt{1+\tau^2} \\
&& b_x(\tau) = 1+\lambda^2(\tau \arctan \tau - \ln\sqrt{1+\tau^2}) \label{bx}
\end{eqnarray}
for the solution of the hydrodynamic expansion differential Eq.~(\ref{eq:Db}),
where we have assumed initial isotropic trapping ($\omega_x=\omega_y=\omega_z\equiv \omega_\text{ho})$, yielding $\lambda = {\tilde \omega}_x/{\tilde \omega}_y=\sqrt{m/m^*}$,  and introduced the dimensionless time $\tau=\omega_\text{ho}t$.

\begin{figure}[tbp]
\centerline{
\includegraphics[width=8.6cm]{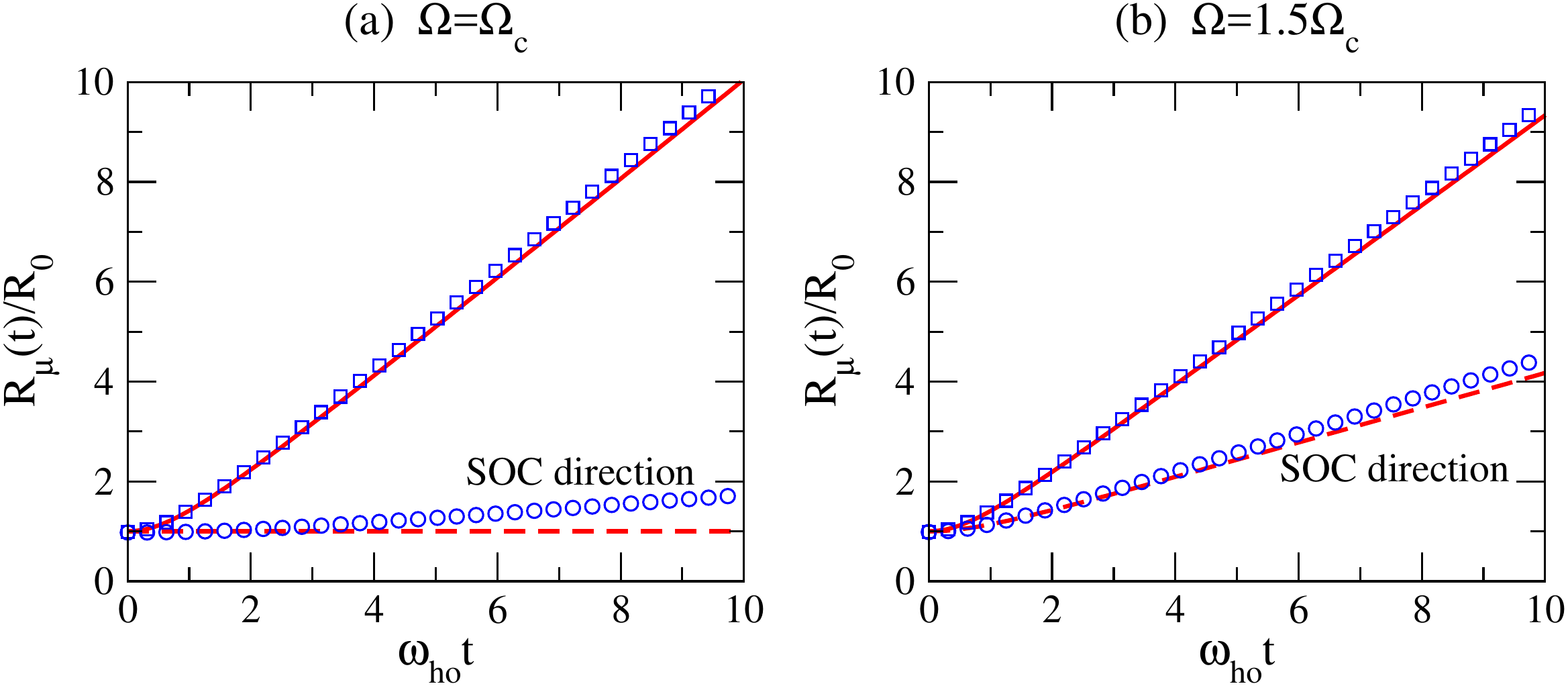}}
\caption{Scaled Thomas-Fermi radius $R_\mu(t)/R_0$ of the expanding spin-orbit coupled Bose-Einstein condensate for different Raman coupling strengths (a) $\Omega=\Omega_\text{c}$ and (b) $\Omega=1.5\Omega_\text{c}$. The dashed red lines and the solid red lines are the results for $R_x(t)$ and $R_{y,z}(t)$ obtained from the hydrodynamic theory. The circles and squares in blue are the corresponding results obtained by numerically solving the Gross-Pitaevskii equations.}
\label{Fig2}
\end{figure}

Using the result $v_x =\frac{\dot{b}_x}{b_x}x$ and Eq. (\ref{bx}) we finally find that, in the limit $m/m^* \to 0$, the spin density can be simplified  to
\begin{equation}
\frac{s_z}{n}= \arctan \tau \sqrt{\frac{\mu_0}{E_\text{r}}} \frac{x}{R_0}
\label{eq:largetimesz}
\end{equation}
with $\arctan \tau \to \pi/2$ for $\tau\to\infty$. In Eq.~(\ref{eq:largetimesz}), we have introduced the initial Thomas-Fermi radius $R_0=R_\mu^{(0)}$ and $\mu_0=m\omega_\text{ho}^2R_0^2/2$. Result (\ref{eq:largetimesz}) shows that, when $\Omega=\Omega_\text{c}$ the above simplified hydrodynamic approach, based on the    condition $s_z\ll n$,  is no longer valid for large expansion times $\tau$, unless $\mu_0$ is sufficiently small compared to $E_\text{r}$.

\begin{figure}[t!]
\centerline{
\includegraphics[width=8.6cm]{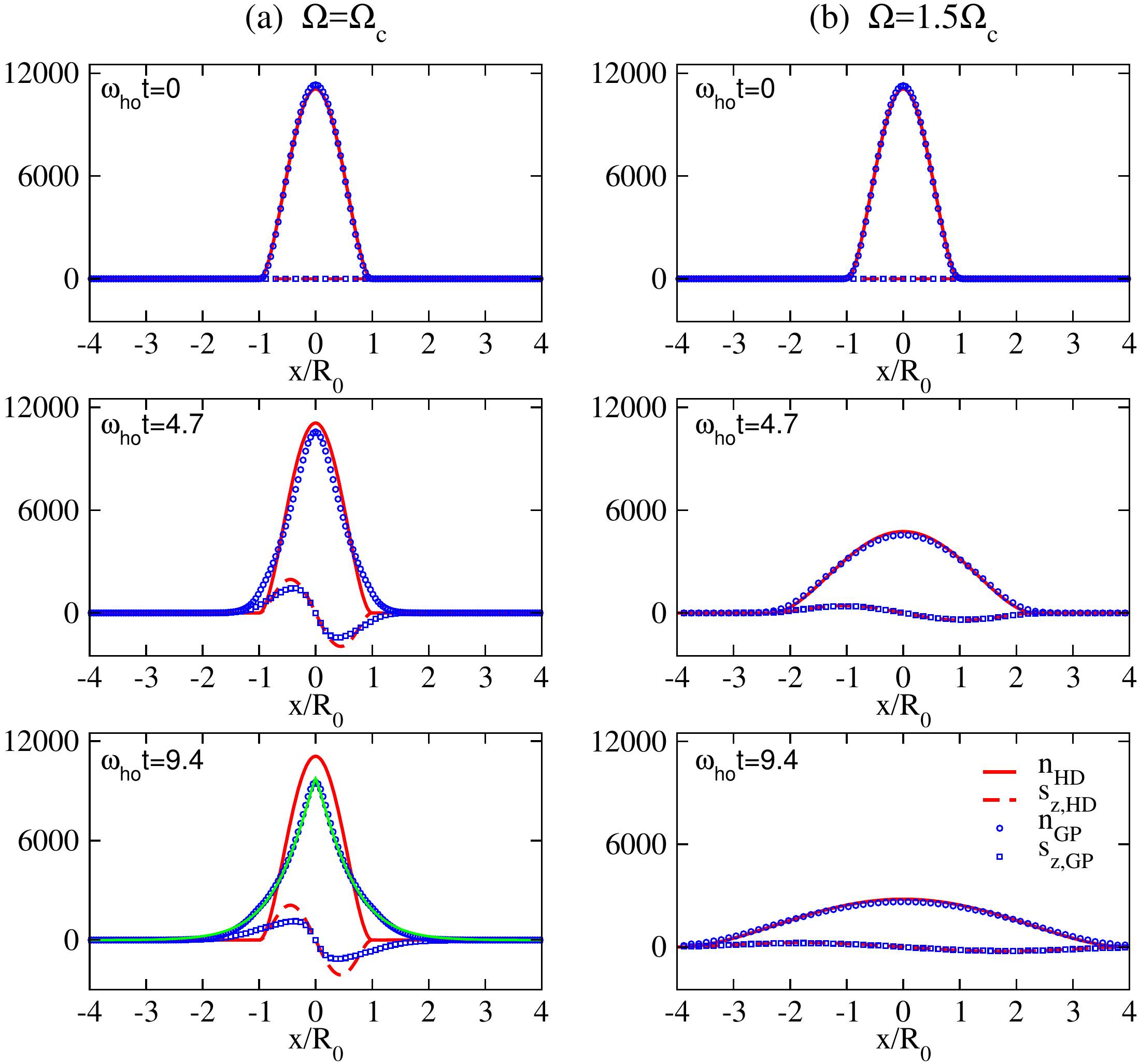}}
\caption{Plot of the total integrated density $n(x)$ and the spin density $s_z(x)$ of the expanding spin-orbit coupled condensate for (a) $\Omega=\Omega_c$ and (b) $\Omega=1.5\Omega_c$. For each value of Raman coupling, the density profiles after an expansion time $\omega_{ho}t=0$, $4.7$, and $9.4$ are shown. The solid and dashed red lines are the results obtained from the hydrodynamic theory while the blue symbols are the results obtained by numerically solving the Gross-Pitaevskii equations, respectively.   In the left-bottom panel, the solid green line is a fiting function of the form $f(x)= \beta \exp(-|x/R_0|^\alpha/\lambda)$ with $\beta=9750$, $\alpha=1.17$ and $\lambda=0.6$.
}
\label{Fig3}
\end{figure}

The hydrodynamic results, based on the solution of the scaling equations for $b_\mu(t)=R_\mu(t)/R_\mu^{(0)}$, are shown as solid and dashed red lines in Fig.~\ref{Fig2} for two different values of the Raman coupling strength $\Omega=\Omega_\text{c} $ and $\Omega=1.5 \Omega_\text{c}$. For the sake of simplicity in these calculations we have assumed initial spherical trapping (similar features are predicted also in the case of anistropic trapping). We find that the expansion along the spin-orbit coupling direction (dashed red lines in Fig.~\ref{Fig2}) is slower than the one along the other directions (solid red lines in Fig.~\ref{Fig2}) and, in particular, it is frozen at the critical transition point between the single-minimum phase and plane-wave phase, exhibiting an effective localization. The corresponding integrated density   $n(x)=\int dydz n(\mathbf{r})=(15N/16R_x(t))(1-x^2/R_x^2(t))^2$ and spin density profiles obtained with the help of Eq.~(\ref{eq:spindensity2}) are instead shown in Fig.~\ref{Fig3} for two different expansion times.

To check the validity of the hydrodynamic approach, a full numerical calculation of the coupled Gross-Pitaevskii equations has been carried out under the same conditions of spherical trapping. We consider a spin-orbit coupled Bose gas with $N=1\times 10^5$ of $^{87}$Rb atoms confined in a spherical harmonic trap with $\omega_\text{ho}=2\pi\times 50$Hz. The Raman laser wavelength $\lambda=782$nm which determines the recoil momentum $k_0=2\pi/\lambda$ and the recoil energy $E_\text{r}$.  For our simulation parameters, we have the initial Thomas-Fermi radius $R_0=8.46\mu m$ and $\mu_0/E_\text{r}\approx 0.2$. After long enough imaginary-time evolution, we prepare the condensate at the ground state for different values of Raman coupling strengths $\Omega$. Then we release the harmonic trap and observe the expansion dynamics of the condensate. The numerical results for the radii are shown by the blue symbols in Fig.~\ref{Fig2} where, in order to compare quantitatively the simulation results with the hydrodynamic predictions, we have numerically calculated the mean square root radii $w_\mu(t)=\sqrt{\langle x_\mu^2\rangle}$ and then determined $R_\mu(t)$ through the Thomas-Fermi relation $R_\mu(t)=\sqrt{7}w_\mu(t)$. The corresponding density and spin density profiles  are instead reported in Fig.~\ref{Fig3}.

The agreement between the simulation and the hydrodynamic theory results is  excellent for Raman coupling $\Omega=1.5\Omega_\text{c}$ at all expansion times. At the critical transition point $\Omega=\Omega_\text{c}$, we observe deviations at large expansion times, reflecting the fact that spin density is not always small compared to the total density. In particular, as shown by the simulation results in Fig.~\ref{Fig2}(a), the expansion of the condensate along the spin-orbit coupling direction at the critical transition point is not completely frozen. Furthermore, as explicitly shown in Fig.~\ref{Fig3}(a), for larger expansion times the total density cannot be longer described as an inverted parabola. This is easy to understand since during the expansion the condensate accumulates a local  phase (and hence a local quasi momentum) which changes the values of the effective mass, resulting in a more complicated behavior of the density distribution. For $\Omega=\Omega_c$, the density at $\omega_\text{ho}t=9.4$ can be approximated by a function of the form $n(x)\approx \beta \exp(-|x/R_0|^\alpha/\lambda)$, reflecting an analogy with the exponential behavior observed in~\cite{Roati2008} during the expansion    in the presence of localization induced by disorder.

\section{Conclusion}
\label{sec:conclusion}

In conclusion, we have studied the expansion dynamics of the spin-orbit coupled Bose-Einstein condensate using a hydrodynamic formalism allowing for analytic scaling solutions. The expansion along the spin-orbit coupling direction is slowed down in a dramatic way near the critical transition point between the single-minimum phase and the plane-wave phase. The slowering is caused by the quenching of the superfluid flow corresponding to a huge increase of the effective mass. The comparison with the  numerical solution of  the coupled Gross-Pitaevskii equations  indicates that our  hydrodynamic approach, which takes into account up to quadratic terms in the velocity field, works very well for the calculation of the density and of the spin density for short expansion times. For a Rashba spin-orbit coupled Bose gas~\cite{Wang2010,Wu2011,Ozawa2012,ZhangYP2012}, the investigation of the expansion dynamics can be carried out similarly using the corresponding hydrodynamic formalism~\cite{Stringari2017}.

\begin{acknowledgements}
We would like to thank F. Dalfovo and P. Engels for discussions and comments. This work was supported by the QUIC grant of the Horizon2020 FET program and by Provincia Autonoma di Trento.
\end{acknowledgements}


\begin{thebibliography}{99}

\bibitem{Anderson1995} M. H. Anderson, J. R. Ensher, M. R. Matthews, C. E. Wieman, and E. A. Cornell, Science \textbf{269}, 5221 (1995).

\bibitem{Davis1995} K. B. Davis, M. -O. Mewes, M. R. Andrews, N. J. van Druten, D. S. Durfee, D. M. Kurn, and W. Ketterle, Phys. Rev. Lett. \textbf{75}, 3969 (1995).

\bibitem{Ernst1998} U. Ernst, A. Marte, F. Schreck, J.Schuster, and G. Rempe, Europhys. Lett. \textbf{41}, 1 (1998).

\bibitem{Stenger1998} J. Stenger, D. M. Stamper-Kurn, M. R. Andrews, A. P. Chikkatur, S. Inouye, H.-J. Miesner, and W. Ketterle, J. Low. Temp. Phys. \textbf{113}, 167 (1998).

\bibitem{Greiner2002} M. Greiner, O. Mandel, T. Esslinger, T. W. H\"ansch, and I. Bloch, Nature \textbf{415}, 39 (2002).

\bibitem{Roati2008} G. Roati, C. D'Errico, L. Fallani, M. Fattori, C. Fort, M. Zaccanti, G. Modugno, M. Modugno, and M. Inguscio, Nature \textbf{453}, 895 (2008).

\bibitem{Roati2004} G. Roati, E. de Mirandes, F. Ferlaino, H. Ott, G. Modugno, and M. Inguscio, Phys. Rev. Lett. \textbf{92}, 230402 (2004).

\bibitem{Pitaevskii2016} L. P. Pitaevskii and S. Stringari, \textit{Bose-Einstein Condensation and Superfluidity}, (Oxford University Press, New York, 2016).

\bibitem{Stringari1996} S. Stringari, Phys. Rev. Lett. \textbf{77}, 2360 (1996).

\bibitem{Castin1996} Y. Castin and R. Dum, Phys. Rev. Lett. \textbf{77}, 5315 (1996).

\bibitem{Kagan1996} Yu. Kagan, E. L. Surkov, and G. V. Shlyapnikov, Phys. Rev. A \textbf{54}, R1753 (1996).

\bibitem{Dalfovo1997} F. Dalfovo, C. Minniti, S. Stringari, and L. P. Pitaevskii, Phys. Lett. A \textbf{227}, 259 (1997).

\bibitem{Lin2011} Y. J. Lin, K. Jimenez-Garcia, and I. B. Spielman, Nature \textbf{471}, 83 (2011).

\bibitem{Fu2011} Z. Fu, P. Wang, S. Chai, L. Huang, and J. Zhang, Phys. Rev. A \textbf{84}, 043609 (2011).

\bibitem{Zhang2012} J.-Y. Zhang, S.-C. Ji. Z. Chen, L. Zhang, Z.-D. Du, B. Yan, G.-S. Pan, B. Zhao, Y.-J. Deng, H. Zhai, S. Chen, and J.-We. Pan, Phys. Rev. Lett. \textbf{109}, 115301 (2012).

\bibitem{Ji2015} S.-C. Ji, L. Zhang, X.-T. Xu, Z. Wu, Y. Deng, S. Chen, and J.-W. Pan, Phys. Rev. Lett. \textbf{114}, 105301 (2015).

\bibitem{Wang2012} P. Wang, Z. Fu, J. Miao, L. Huang, S. Chai, H. Zhai, and J. Zhang, Phys. Rev. Lett. \textbf{109}, 095301 (2012).

\bibitem{Cheuk2012} L. W. Cheuk, et al. Phys. Rev. Lett. \textbf{109}, 095302 (2012).

\bibitem{Qu2013} C. Qu, C. Hamner, M. Gong, C. Zhang, and P. Engels, Phys. Rev. A \textbf{88}, 021604 (2013).

\bibitem{Olson2014} A. J. Olson, S.-J. Wang, R. J. Niffenegger, C.-H. Li, C. H. Greene, and Y. P. Chen, Phys. Rev. A \textbf{90}, 013616 (2014).

\bibitem{Zhai2015} H. Zhai, Rep. Prog. in Physics, \textbf{78}, 026001 (2015).

\bibitem{Li2015} Y. Li, G. I. Martone, and S. Stringari, Annual Review of Cold Atoms and Molecules, World Scientific, \textbf{3}, 201 (2015).

\bibitem{Zhu2012} Q. Zhu, C. Zhang and B. Wu, Euro. Phys. Lett. \textbf{100}, 50003 (2012).

\bibitem{Zheng2013} W. Zheng, Z.-Q. Yu, X. Cui, and H. Zhai, J. Phys. B \textbf{46}, 134007 (2013).

\bibitem{Ozawa2012a} T. Ozawa, L. P. Pitaevskii, and S. Stringari, Phys. Rev. A \textbf{87}, 062610 (2012).

\bibitem{Martone2012} G. I. Martone, Y. Li, L. P. Pitaevskii, and S. Stringari, Phys. Rev. A \textbf{86}, 063621 (2012).

\bibitem{Zhang2016} Y.-C. Zhang, Z.-Q. Yu, T. K. Ng, S. Zhang, L. Pitaevskii, and S. Stringari, Phys. Rev. A \textbf{94}, 033635 (2016).

\bibitem{Stringari2017} S. Stringari, arXiv:1609.04694, to appear on Phys. Rev. Lett. (2017).

\bibitem{Khamehchi2016} M. A. Khamehchi, Khalid Hossain, M. E. Mossman, Y. Zhang, Th. Busch, M. M. Forbes, and P. Engels, arXiv:1612.04055, to appear on Phys. Rev. Lett. (2017).

\bibitem{Ho2011} T.-L. Ho and S. Zhang, Phys. Rev. Lett. \textbf{107}, 150403 (2011).

\bibitem{Li2012} Y. Li, L. P. Pitaevskii, and S. Stringari, \textbf{108}, 225301 (2012).

\bibitem{Wang2010} C. Wang, C. Gao, C.-M. Jian, and H. Zhai, Phys. Rev. Lett. \textbf{105}, 160403 (2010).

\bibitem{Wu2011} C. Wu, I. Mondragon-Shem, and X.-F. Zhou, Chin. Phys. Lett. \textbf{28}, 097102 (2011).

\bibitem{Ozawa2012} T. Ozawa and G. Baym, Phys. Rev. Lett. \textbf{109}, 025301 (2012).

\bibitem{ZhangYP2012} Y. Zhang, L. Mao, and C. Zhang, \textbf{108}, 035302 (2012).


\end{thebibliography}
\end{document}